\documentclass[]{article}

\usepackage[unicode,psdextra,hypertexnames=false]{hyperref}
\usepackage{enumitem}
\usepackage{amssymb}
\usepackage{amsmath}
\usepackage[linesnumbered,ruled,vlined]{algorithm2e}
\usepackage{float}
\usepackage{hyperref}
\usepackage{graphicx}

\usepackage{xcolor}
\usepackage{soul}

\newcommand\pdfmath[1]{\texorpdfstring{$#1$}{#1}}

\newtheorem{definition}{Definition}

\title{Differential Privacy: What is all the noise about?}
\date{}

\author{Roxana Danger\footnote{regulaition.com}}

\begin{document}

\maketitle
\begin{sloppypar}

\begin{abstract}
Differential Privacy (DP) is a formal definition of privacy that provides rigorous guarantees against risks of privacy breaches during data processing. It makes no assumptions about the knowledge or computational power of adversaries, and provides an interpretable, quantifiable and composable formalism. DP has been actively researched during the last 15 years, but it is still hard to master for many Machine Learning (ML)) practitioners. This paper aims to provide an overview of the most important ideas, concepts and uses of DP in ML, with special focus on its intersection with Federated Learning (FL).
\end{abstract}

\section{Introduction}

Privacy was first defined by Alan F. Westin as ``the claim of individuals, groups, or institutions to determine for themselves when, how, and to what extent information about them is communicated to others.''~\cite{Westin1967}. The rise of databases, internet, social networks, location technology, digital searching and advertising, along with all their tracking mechanisms, has made digital privacy a huge concern, to the point where governments and international institutions have dictated laws and regulations to provide minimal privacy guarantees for individuals.

Researchers have tackled the problem using techniques as encryption, anonymity, obfuscation, diversity, delta-presence, and strict security measures to reduce access to the data, but all of these have proven to be insufficient from security or usability perspectives~\cite{Jian2016, 10.1007/978-3-030-27813-7_5}. Differential privacy (DP) was born in this context to bring a new way of measuring and guaranteeing privacy. 
DP is a conceptual framework with a rigorous mathematical foundation that can be applied to any type of algorithm, and it sits at the intersection of computer science, statistics, economics, law and ethics. It has gained an enormous support by the research community, and some companies and governments are beginning to incorporate it into their own data pipelines~\cite{Hawes2020, bird2020, Greig2019, Google2019, Google2020, DBLP:journals/corr/JohnsonNS17_uber, Hutchinson2020_FB, Herdagdelen2020_FB, Diethe2020_AMZ, Feyisetan2020_AMZ, Rogers20202_LinkedIn, Pihur2018}. 




Unlike previous privacy definitions, DP is not based on assumptions about a privacy attacker (whose characterization is a very difficult problem in itself), but rather on the notion on the concept of \emph{privacy leakage}~\cite{dwork2006} (also referred as \textit{privacy loss}). Data processors using DP determine the level of privacy leakage users could tolerate or is ethically acceptable, and applied DP mechanisms to guarantee these limits are not exceeded. 

Many different interrelated concepts are associated with DP: privacy loss, mechanisms of DP, local and centralized DP, uses of DP in statistics, Machine Learning and Federated Learning. This paper aims to provide an overview of the most important ideas, concepts and uses of DP in ML, with special focus on its intersection with Federated Learning, as well as to provide a practical guideline for applying DP in a concrete problem.

Privacy leakage is usually denoted by $\epsilon$; it is a provable limit to the inference that can be derived about an individual private, sensitive, information from observing the outcome of a DP-mechanism. This formalism is often called $\epsilon$-DP and its strong mathematical properties makes any $\epsilon$-differentially private mechanism a trustful one \cite{reusche2021priorfree}. The original definitions of $\epsilon$-DP and privacy loss as well as the properties associated to them were introduced by Cynthia Dwork and her colleagues~\cite{dwork2006} and they are discussed in Section~\ref{sec:epsDP}. 

The main idea behind DP is that by adding certain calibrated noise to the outcome of a non-DP algorithm (called also mechanism in this context), it converts to an equivalent DP one. The noise is calibrated based on the maximum privacy loss guarantee while maximizing the utility of the DP results. The noise is sampled from statistical distributions like Laplace, gaussian, geometrical- thus producing Laplace, gaussian, geometrical DP-mechanisms. These basic mechanisms are described in Section~\ref{sec:dpmech}. More complex mechanisms have been proposed to fulfil requirements of privacy in multiple scenarios. A brilliant taxonomy of all these variants with formal definitions, axioms and theorems that relate to them can be found at: \cite{desfontaines2020sok}, but in Section~\ref{sec:DPvariants} the interested reader can peruse a brief summary of them.

When solving a ML task (i.e. classification, regression, clustering), DP can be applied at any phase of the ML pipeline (preprocessing, parameter optimization, loss computation, etc.) using the mechanisms previously described. Particularly important are the extensions of ML optimization algorithm SGD and the ensemble techniques to convert them into DP mechanisms, given way to DP-SDG~\cite{abadi2016deep} and the ensemble-DP framework~\cite{mivule2012towards}, which have led to PATE~\cite{papernot2017semisupervised}, a well recognised DP framework in ML. Section~\ref{sec:ensemble-DP-and-DPSGD} is devoted to describe these algorithms.

A second path to obtain privacy in ML is through the use of synthetic data
instead of real data. Generative Adversarial Networks (GAN) that have proved to be powerful tool for generating synthetic data. So, Section~\ref{sec:DP-GAN-and-PATE-GAN} introduces some works that combines DP with GAN models. 

Strong guarantees for privacy can be obtained by using the above methods. However, the original private data are still saved and maintained in a server, and the users have to trust the data holder about their data. In some scenarios, 
private data is not stored in a central database and the user can add noise to the data before allowing any learning task to be performed on it. This setting is known as \textit{local differential privacy} (LDP) and the most common use cases are discussed in Section~\ref{sec:localdp}.

Finally, another way of reducing the problems associated with data privacy in ML (and at the same time improve the generalization power of models) is using \textit{Federated Learning (FL)}. Federated Learning, in contrast with traditional centralised ML, trains an algorithm across multiple decentralized servers holding local data, without exchanging or downloading the data locally. The core element of FL is that data never leave their original location, therefore privacy is somehow guaranteed. Other types of attacks in the FL setting are still possible and multiple counter-attack solutions have been proposed, as described in Section~\ref{sec:dpFL}. 

In Section~\ref{sec:discussion} we summarise the themes discussed in this study and give some guidelines for applying DP in practice. Finally, Section~\ref{sec:conclusions} contains the conclusions of the paper.

\section{ \pdfmath{\epsilon}-differential privacy definition} \label{sec:epsDP}

Differential privacy 
is defined on the concept of adjacent datasets. Two datasets are adjacent if one of them can be created by removing (or changing) a single example from the other dataset. DP is formalized as follows~\cite{dwork2006}:

\begin{definition}[$\epsilon$-DP] 
	A randomized mechanism of function $\mathcal{M}: \mathcal{D} \rightarrow \mathcal{R}$ with domain $\mathcal{D}$ and range $\mathcal{R}$ satisfies $\epsilon$-differential privacy if, for any two adjacent inputs $d, d' \in \mathcal{D}$ and for any subset of outputs $\mathcal{X} \subseteq \mathcal{R}$, it holds that:
	
	\[ Pr[\mathcal{M}(d) \in X] \leq e^\epsilon Pr[\mathcal{M}(d') \in X] \].
	
\end{definition}

$Pr[\mathcal{M}(d) \in X]$ is the probability that when the mechanism $\mathcal{M}$ runs in the dataset $d$ its output is in the set $X$. So, this equation says that the change in likelihood of the outcome of $\mathcal{M}$ by using $d$ or $d'$ is bounded by $e^\epsilon$. 
If the likelihood of obtaining a response is as much tripled when any row of the dataset is removed or added, $\frac{Pr[\mathcal{M}(d) \in X]}{Pr[\mathcal{M}(d') \in X]} = 3$, we are in presence of a $1.1$-DP mechanism, as $e^\epsilon=3$ when $\epsilon=1.1$

The definition of adjacent datasets depend on the application, but it allows the estimation of a value of the \textit{privacy loss}, $\epsilon$, when a mechanism is applied to the adjacent datasets, that is, how much an observer can learn about a single example by applying the mechanism to a dataset. 
In fact, using the Bayes's Rule and the above definition, it can establish the lower and bounds of the posterior probability, after applying a $\epsilon$-DP mechanism and it can be used as an indication of the knowledge gain an observer of $M$ can obtain. Specifically, given a prior probability $p$ representing the uncertainty of the current knowledge $K$ about an example, the uncertainty about the example, after applying the mechanism $\mathcal{M}$, i.e. the posterior probability, $P(\mathcal{M}(d) \in X | K)$, can be bounded as:

\[
 \frac{p}{e^\epsilon + (1-e^\epsilon)*p} \le Pr(\mathcal{M}(d) \in X | K) \le \frac{e^\epsilon*p}{1+(e^\epsilon-1)*p} 
\]

\begin{figure}[ht]
	\centering
	\centering
	\begin{minipage}[b]{0.45\textwidth}
		\includegraphics[width=\textwidth]{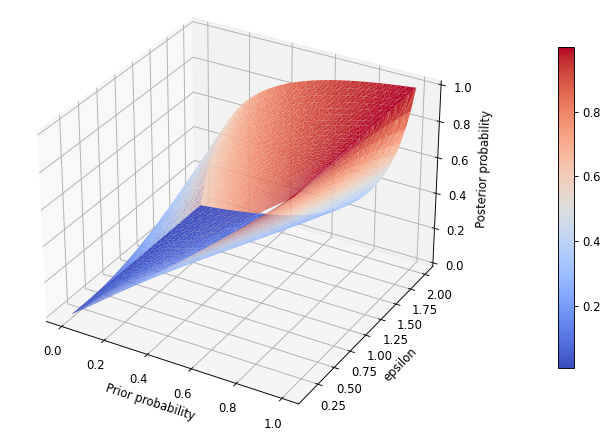}
		\caption{(a)}
	\end{minipage}
	\hfill
	\begin{minipage}[b]{0.45\textwidth}
		\includegraphics[width=\textwidth]{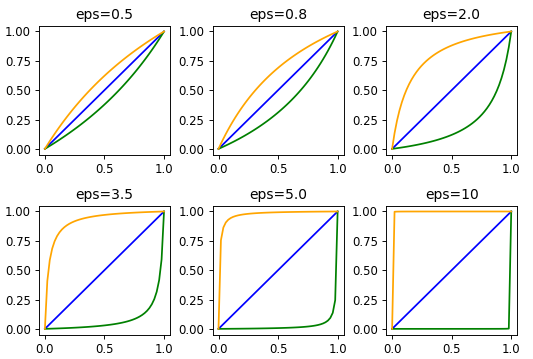}
		\caption{(b)}
	\end{minipage}

	\caption{(a) Upper and lower bounds of posterior probability based on epsilon and a prior probability; (b)Upper and lower bounds breakdown at different epsilon values.}\label{fig:bounds}
\end{figure}

Figure~\ref{fig:bounds} shows these bounds graphically. The greater $\epsilon$, the greater the posterior probability, therefore, the greater the privacy loss. Therefore, choosing smaller $\epsilon$ guarantees a more privacy-preserving (with less privacy loss or leakage) mechanism. 


The next questions are: given a non-DP mechanism, how can it be made differentially private? And which value of $\epsilon$ should be chosen? 

The answer to the first question is to add uncertainty - a probabilistic noise- to the outcome of the mechanism. This noise addition provides the necessary randomness to protect sensitive user data. It was proved in~\cite{dwork2006}, that a mechanism $\mathcal{M}$ can be made $\epsilon$-differentially private by adding an independent Laplace noise vector $v$:

\[ Pr[v] = Lap(x | \mu=0, b) = \frac{1}{2b} e^\frac{|| v ||_1}{b} \varpropto e^\frac{||v||_1}{b} \]

where: $||.||_1$ is the $L_1$ norm, $b=S(\mathcal{M})/\epsilon$ is defined in terms of the \textit{sensitivity of the mechanism} $\mathcal{M}$, $S(\mathcal{M})$, defined by:

\[ S(\mathcal{M}) = max_{d, d'} ||\mathcal{M}(d) - \mathcal{M}(d')||_1 \]

Informally, the sensibility of a mechanism expresses the maximum expected change of its response considering any pair of adjacent datasets. 

A `good' $\epsilon$ value should be low and at the same time maintain a certain level of utility of the answers provided by the DP mechanism. In other words, it should still produce responses near to the true values. Its estimation depends on the application, and it is a challenging open question to establish the right values for problems in different scenarios~\cite{dwork2014}.

This definition assumes that privacy is only lost due to the observability of the mechanism. A natural extension is assuming that there are also some uncontrolled events that can affect privacy. This notion of uncontrolled risk is captured in the definition of $(\epsilon, \delta)$-DP as follows~\cite{dwork2006our}:

\begin{definition}[($\epsilon, \delta$)-DP] 
	A randomized mechanism of function $\mathcal{M}: \mathcal{D} \rightarrow \mathcal{R}$ with domain $\mathcal{D}$ and range $\mathcal{R}$ satisfies $(\epsilon, \delta)$-differential privacy if any two adjacent inputs $d, d' \in \mathcal{D}$ and, for any subset of outputs $\mathcal{X} \subseteq \mathcal{R}$, it holds that:
	
	\[ Pr[M(d) \in X] \leq e^\epsilon Pr[M(d') \in X] + \delta \].
	
\end{definition}

In this case, $\delta$ is a variable representing the probability that DP is broken by factors external to the DP mechanism. As for $\epsilon$, we would like to set $\delta$ to a very low value. To avoid the worst-case scenario of always violating privacy of a $\delta$ fraction of the dataset, the standard recommendation is to choose $\delta \ll 1/N$ or even $\delta=negl(1/N)$, where $N$ is the number of contributors. This strategy avoids the possibility of one particularly devastating outcome, but other forms of information leakage may remain \cite{mironov2017}.

\subsection{Features of differential privacy}

In addition to permit attack modelling, the above definition of DP allows to quantify privacy loss and satisfy multiple mathematical features:

\begin{description}[labelsep=0pt]
	\item [Composability]: Differential privacy composability states that the joint distribution of differentially private mechanisms are also differentially private \cite{mcsherry2009privacy}:
	\begin{itemize}
		\item Sequential: If there are $n$ independent mechanisms: $\mathcal{M}_1, \dots , \mathcal{M}_n$, whose privacy guarantees are $\epsilon_1, \dots, \epsilon_{n}$-differential privacy, respectively, then any function $g(\mathcal {M}_1, \dots, \mathcal{M}_n)$ is $\sum \limits_{i=1}^{n} \epsilon_i$-differentially private.
		\item Parallel: If the previous mechanisms are computed on disjoint subsets of the database then the function $g$ would be $(\max \limits_{i} \epsilon_i)$-differentially private.
	\end{itemize}

	This means that we can combine DP-mechanisms and still estimate the privacy loss of the combined mechanisms.

	\item [Robustness to post-processing]: Any deterministic or randomized function $f$ defined over the image of a $(\epsilon, \delta)$-differentially private mechanism $\mathcal{M}$ is also differentially private, that is $f({\mathcal{M}})$ is also a $(\epsilon, \delta)$-differentially private mechanism. 
	
	This means that we can use the output of a DP mechanism in any other process and the final result will not reveal more than what was already revealed by the DP-mechanism.
	
	\item [Group privacy]: According to \cite{dwork2006differential}, if we would like to protect databases differing in $c$ rows (that is, avoiding that an attacker with auxiliary information can deduce if the dataset contains the information of $c$ participants), we only need to bound the privacy value as $\epsilon c$ instead of $\epsilon$, in this case, for $d_1$, $d_2$ differing on $c$ items:
	
	\[ Pr[M(d_1) \in X] \leq e^{\epsilon c} Pr[M(d_2) \in X] + \delta \]
	
	So, every group of $c$ items is $\epsilon$-differentially private protected and each item is $\frac{\epsilon}{c}$-differentially private protected.
	
\end{description}

These features combined mean DP is a comprehensive framework for privacy whose tools allows: 1)  measurement of the privacy guarantee that a system provides and 2) understanding of its interaction with other DP and non-DP mechanisms. Intense research has been directed towards defining other types of DP mechanisms and/or describing variants that consider other elements besides $epsilon$, to define more realistic or useful DP-mechanisms.


\section{Differential privacy mechanisms}\label{sec:dpmech}

We have previously described Laplace mechanisms, but there are other probability distribution functions that have been proved useful as differential privacy mechanisms, such as:

\begin{itemize}
	\item \textit{Gaussian mechanism}: Very similar to Laplace mechanism, but it adds Gaussian noise instead of Laplacian noise, and it needs parameter $\delta>0$ (which is an acceptable assumption). Although the general accuracy of gaussian mechanism is lower than the Laplacian, it has gained great attention, for example in ML applications, since L1 or L2 norm are both acceptable sensibility functions, and the noise it produces at the point of the true value is lower than in the Laplace case, which improves the accuracy of the mechanism if the sensibility is low~\cite{dwork2006differential, holohan2021secure} 
	
	\item \textit{Geometric mechanism}: This mechanism draws noise from the double-geometric distribution, which is the discrete version of the Laplace distribution. It only protects count queries, but the specialization improves performance over the more general Laplace mechanism~\cite{1536414.1536464} 
	
	\item \textit{Uniform mechanism}: The uniform noise distribution is near optimal and it can be seen as a special case of the bounded Laplace mechanism in a regime with $\epsilon=0$. \cite{geng2013, geng2019privacy}
	
	\item \textit{Binary mechanism}: Given a binary input value, the mechanism randomly decides to flip the binary value or not, in order to satisfy differential privacy \cite{holohan2017}.
	
	\item \textit{Exponential mechanism}: Achieves differential privacy on numerical and categorical inputs, by randomly choosing an output value for a given input value, with greater probability assigned to the values that are closer to the input, as measured by a given utility function \cite{mcsherry2007}.
	
	\item \textit{Bingham mechanism}: Uses Bingham distribution for estimating the first eigenvector of a covariance matrix \cite{kanti2018}.

\end{itemize}


A large survey of applications of DP in multiple industry setup scenarios can be seen at \cite{DBLP:journals/corr/abs-1812-02282}.
and notable implementations of the above mechanisms (and the variations explained in~Section~\ref{sec:DPvariants}) are available at:
\begin{itemize}
	\item https://github.com/IBM/differential-privacy-library
	\item https://github.com/google/differential-privacy
	\item https://github.com/yuxiangw/autodp, and 
	\item https://github.com/opendp/smartnoise-core
	\item https://github.com/tensorflow/privacy
\end{itemize}

DP is not hard to use, and the data holder can maintain the original data but share analysis over these data with a certain privacy guarantee, that is controlled according to the DP parameters. However, the DP results have a reduced utility with respect to using a non DP procedure. Selecting the right parameters to compensate the trade-off between privacy and accuracy is a tough problem. Finally, due to the dimensionality curse, the more data needs to be freed in a DP fashion, the more difficult this trade-off become. Therefore, a lot of research has been done to define mechanisms or scenarios whose conditions allow to identify tighter privacy loss bounds.  

\section{Other variants of Differential privacy}\label{sec:DPvariants}

There are more than 200 variants and extensions of DP definitions in the literature, each of them focused on different scenarios, datatypes and attacker models. A brilliant taxonomy of these variants with formal definitions, axioms and theorems that relate to them can be found at: \cite{desfontaines2020sok}. In this work the DP definitions are partitioned into the following seven categories depending on which features of the original definition were modified:

\begin{description}[labelsep=0pt]
	\item[Quantification of privacy loss]: Whilst the original definition of privacy loss (that considers only the parameter $\epsilon$) models the risk of the worst-case scenario, additions such as considering a small probability error $\delta$ are more natural and offer better composition properties. Particular attention has been given to averaging the privacy loss (using other measures of divergence obtained by the privacy mechanism, such as Kullback-Leibler or the Renyi-divergence \cite{mironov2017} \footnote{Renyi-DP strong interpretation along with the well-suited composition of heterogeneous mechanisms have made Renyi-DP one of the most popular frameworks for privacy. See \url{https://www.johndcook.com/blog/2018/11/21/renyi-differential-privacy/} for a concise description of the Renyi-DP.}). It is also worth mentioning the case of controlling the privacy loss of the tail of the distributions, in which both average and rare cases are private-preserved.
	
	\item[Neighbourhood definition]: these variants are based on changing the definition of sensitivity, e.g. by increasing or reducing the number of rows that two adjacent datasets can differ on, adding a policy that specifies which records are sensitive, or applying the definition to a particular datatype. 
	
	\item[Variance of the privacy loss]: in DP $\epsilon$ is applied uniformly, but in practice some users require more protection than others. Some definitions allow the privacy loss to vary across users either explicitly (using an user acceptance level) or implicitly (by averaging the risk or leaving some users at risk).
	
	\item[Background knowledge]: normally, it is assumed the attacker to have full knowledge of the dataset, and their only uncertainty is if a user belongs to the dataset or not. Instead of using adjacent datasets, these types of DP consider the existence of certain probability distributions on the data that is used to simulate the attacker's knowledge and restrict the privacy guarantees. This is an active area of research, but not currently used in practice as there is no guarantee about the limited attacker background knowledge.
	
	\item[Change in formalism]: other statistical formalisms have been proposed, for example using hypothesis testing where DP is interpreted as the probability of an attacker to know that a particular outcome comes from a dataset or its neighbourhood dataset. This type of formalism have evolved on the \textit{f-differential privacy}, that guarantees the composition of any mechanisms and provides an exact formula for composition, therefore gaining a lot of practical attraction in the SOTA DP-libraries.
	
	\item[Relativization of knowledge gain]: beyond the probabilistic information bounded by DP, information can be leaked in other ways. These other types of information leakage can be modelled using auxiliary functions or algorithms.
	
	\item[Computational power]: these variants consider that the computational power an adversary has can lead to a different privacy guarantee level. Limited computational power is a reasonable assumption, but for a large class of queries or algorithms, it cannot provide significant benefits over classical DP in a typical client-server setup, thus, existing works are focused on federated settings \cite{10.14778/3291264.3291274}.
	
\end{description}

Some works extend the concept of DP based on a mix of the above categories or scenarios. They are also discussed in \cite{desfontaines2020sok}.
 
\section{DP techniques in ML}\label{sec:dpML}

Thanks to the post-processing feature of DP, we can apply DP techniques at various stages of the ML process and still achieve a differentially private final model. DP can be applied to: 1) the datasets during data preprocessing, 2) optimization algorithm whilst training the model parameters, 3) the result of the loss function just before updating parameters or 4) the final trained parameters. Surveys about DP applied to classical ML algorithms (such as Naive Bayes, Linear Regression, classification, Support Vector Machines, dimensionality reduction and time series) can be found in \cite{ji2014differential}, and \cite{sarwate2013signal}. 

In this section we will focus on general methods that can be used in a broader list of ML algorithms instead of a particular one. In Section~\ref{sec:ensemble-DP-and-DPSGD}, the extensions of ML optimization algorithm stochastic gradient descent (SGD) and the ensemble techniques to convert them into DP mechanisms, given way to DP-SGD and the ensemble-DP framework, which have led to Private Aggregation of Teacher Ensembles (PATE), a well recognised DP framework in ML. Then, in Section~\ref{sec:DP-GAN-and-PATE-GAN} we introduce a second path to obtain privacy in ML, through the use of synthetic data instead of real data and its solution by using Generative Adversarial Networks (GAN) and combining it with the PATE algorithm (PATE-GAN).

\subsection{Ensemble-DP and DP-SGD}\label{sec:ensemble-DP-and-DPSGD}

One of the first works towards this direction appears in \cite{mivule2012towards}. In this paper, the authors propose an \textbf{\textit{ensemble-DP framework}}. It works by first applying a strong privacy-preserving mechanism to the dataset and then using an ensemble classifier on top of the perturbed data. It introduces an association between data utility and  increasing the number of weak decision tree learners, where a combined adjustment of the privacy parameters and an increase in the number of weak learners in the ensemble compensated for the increase in classification error after application of DP. Multiple DP-ML applications rely on this framework, due to its simplicity. 

Since most ML algorithms optimize an objective function that minimizes the learning error using stochastic gradient descendent (SGD), and given that SGD is inherently a random mechanism, creating a differentially private SGD is a good foundation for the task. \textbf{\textit{DP-SGD}} was formulated in~\cite{abadi2016deep} and we have reproduced in Algorithm~\ref{alg:dp-sdg}. As with SGD, the algorithm begins randomly assigning parameters $\theta$ of the model and then iteratively updating them according to the loss function $\mathcal{L}(\theta)$. The difference is that in the DP-SGD, at each iteration the gradient is computed for a random subset of examples, then clipped with the L2 norm of the gradients and a gaussian noise is added to protect privacy. The number of iterations, $T$, satisfies  $T \gg N/B$, where $N$ is the number of examples and $B$ is the mini-batch size, to guarantee each sample is used multiple times during the training, .

\noindent
\begin{algorithm}
\caption{DP-SGD: Differentially-private stochastic gradient descent}\label{alg:dp-sdg}

\KwIn{
\begin{itemize}\itemsep0em 
	\item Training data: $\mathcal{D} = \{\mathbf{x}_1,\dots, \mathbf{x}_N\}$
	\item Loss function: $\mathcal{L}[\theta] = \frac{1}{N}\sum_i\mathcal{L}(\theta,\mathbf{x}_i)$
	\item learning rate; $\eta_t$
	\item noise scale: $\sigma$
	\item batch size: $B$
	\item gradient norm bound: $C$
\end{itemize}
}
\KwOut{Final parameters: $\theta_T$}
\vspace{0.3cm}
Initialize $\theta_0$ randomly

\For {$t=1,\dots,T$}{

Randomly sample a batch $\mathcal{D}_t \subset \mathcal{D}$ with sampling probability $B/N$

Compute the gradient $\mathbf{g}_t[\mathbf{x}_i] = \nabla_{\theta} \mathcal{L}[\theta_{t-1},\mathbf{x}_i]$ for each $\mathbf{x}_i \in \mathcal{D}_t$

Clip the gradients:
\begin{equation}
	\mathbf{g}_t[\mathbf{x}_i] \leftarrow \frac{\mathbf{g}_t[\mathbf{x}_i]}{\max\left[1, \frac{\Vert \mathbf{g}_t(\mathbf{x}_i) \Vert}{C}\right]}
\end{equation}

Add noise $\xi_{i,t}\sim \mbox{Norm}_{\xi_{i,t}}[\mathbf{0},\sigma^{2}C^{2}\mathbf{I}]$
\begin{equation}
	\mathbf{g}_t[\mathbf{x}_i] \leftarrow \mathbf{g}_t[\mathbf{x}_{i}] + \xi_{i,t}
\end{equation}

Average the clipped, noisy gradients
\begin{equation}
	\bar{\mathbf{g}}_t = \frac{1}{B} \sum_{x_i\in{D}_{t}} \mathbf{g}_t[\mathbf{x}_i]
\end{equation}

Take a step in the gradient direction
\begin{equation}
	\theta_t \leftarrow \theta_{t-1} - \eta_t\bar{\mathbf{g}}_t
\end{equation}
}

The overall privacy cost $(\epsilon, \delta)$ can be computed by using a privacy accounting method
\end{algorithm}

Notice that the algorithm is expressed in function of a noise scale, and the value of $\epsilon$ is computed by using the \textit{moment accountant method} that was described in the same paper to keep a better track of the privacy loss. Besides, for a fixed level of differential privacy, there is a connection between the amount of noise $\sigma$ and the amount of clipping $C$. A larger value of $C$ does less clipping, but requires more noise to retain the same degree of privacy (compensating the effect of sensitive values). Specifically, it was proved that the Algorithm~\ref{alg:dp-sdg} is $(\epsilon, \delta)$-DP when using $\sigma = \Omega(q \frac{\sqrt{T log(1/\delta)log(T/\delta)}}{\epsilon})$, being $q=B/N$, sampling probability.


An important outcome of this paper is that in their experiments, DP-SGD improved generalization performance and that the gap between training and testing accuracy tended to be smaller, particularly for small privacy budgets.

\begin{figure}[ht]
	\centering
	\includegraphics[width=\textwidth]{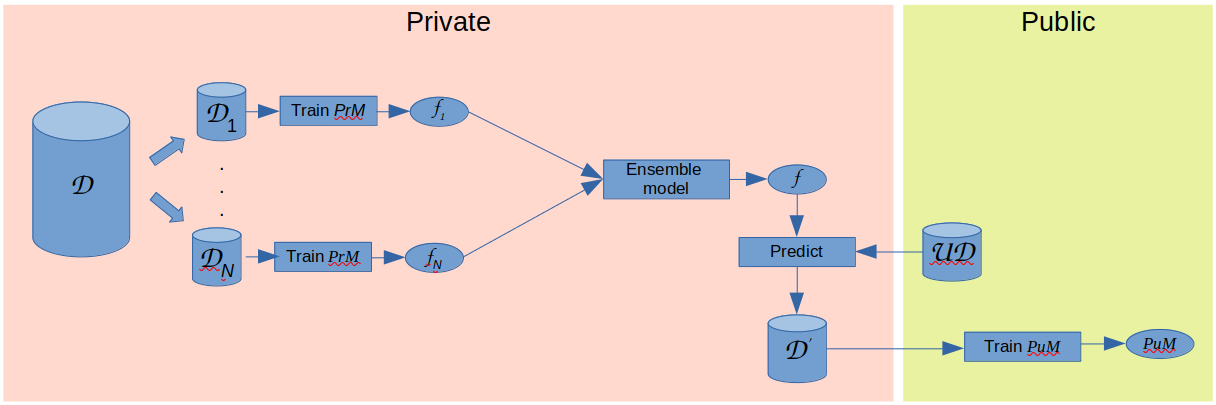}	
	\caption{General architecture and data processing of PATE algorithm}\label{fig:pate}
\end{figure}

Another paper, \cite{papernot2017semisupervised}, mixes the ideas in the ensemble-DP framework and DP-SG in a proposal known as the \textbf{PATE} algorithm. The PATE framework basically allows a public model to learn by noisily ensembling the predictions of multiple models. In this setup it is assumed that there is a private dataset $\mathcal{D}$, a private model $PrM$ and a public unlabelled dataset $\mathcal{UD}$. The private dataset is split in multiple parts to produce multiple teacher models $f_i$. An ensemble model, called aggregate teacher $f$, is defined as noisy voting of the teacher models and then, used to produce a private labelled dataset $\mathcal{D'}$ from the public unlabelled one. $\mathcal{D'}$ is then used to train a public model that is called the student model. Notice that 1) the labels produced by the aggregate teacher are never revealed, nor are the parameters of the teacher models; 2) aggregate model as well as the labels are differentially private thanks to the added noise, but the privacy budget of the model is limited and possibly expended during the predictions; 3) the produced differentially private dataset can now be used to generate any student model that is also differentially private by the post-processing property of DP. A general overview of the whole data process in PATE Algorithm is described in~Figure\ref{fig:pate}. Details are provided in Algorithm\ref{alg:pate}. PATE\footnote{Code available at: \url{https://github.com/tensorflow/privacy/tree/master/research/pate_2018}.} and the following improvements in \cite{papernot2018scalable} have shown that this framework produces higher test set accuracy with lower privacy budgets. 

\noindent
\begin{algorithm}
	\caption{PATE: Private Aggregation of Teacher Ensembles}\label{alg:pate}
	
	\KwIn{
		\begin{itemize}\itemsep0em 
			\item Training data: $\mathcal{D}$
			\item Number of splits: $N$
			\item Private model: $PrM$
			\item Public model: $PuM$
			\item Public unlabelled data: $\mathcal{UD}$
		\end{itemize}
	}
	\KwOut{Final parameters: $\theta_T$ of $PuM$}
	\vspace{0.3cm}
	Split randomly the dataset $\mathcal{D}$ in $N$ parts, $\mathcal{D}=\{\mathcal{D}_1,\dots, \mathcal{D}_N\}$ 
	
	\For {$i=1,\dots,N$}{
		Generate $f_i$ by training $PrM$ with the subset $\mathcal{D}_i$
	}

	Generate the aggregate teacher model, a noisy ensemble voting model: 
	$f[x] = \underset{j}{\operatorname{argmax}} \left[ \sum_{i} [f_i[x]=j] + \xi \right]$, with $\xi \sim Lap[\epsilon^-1]$
	
	Produce $\mathcal{D'}$ a private labelled dataset by using $f$ to label $\mathcal{UD}$
	
	Use $\mathcal{D'}$ to train $PuM$. Let $\theta_T$ be the trained parameters of $PuM$
	
\end{algorithm}

This algorithm showed another direction in private ML; instead of creating differentially private models, we could produce differentially private datasets that can be used to train differentially private models. 

Yet another direction in private ML is model creation based trained with \textit{synthetic data} instead of using real data. Synthetic data generation systems aim at mirroring the statistical distributions of the original data without revealing information of any particular data instance. But, generating private synthetic data is known to be hard in the worst case \url{https://eccc.weizmann.ac.il/report/2010/017/} and therefore it is necessary to use techniques such as Generative Adversarial Network (GAN), a type of NN designed to simulate (or recreate) noisy examples that follow the patterns of an input dataset. This is the topic of the following section.

\subsection{DP GAN models: DP-GAN and PATE-GAN}\label{sec:DP-GAN-and-PATE-GAN}

\begin{figure}[ht]
	\centering
	\includegraphics[width=\textwidth]{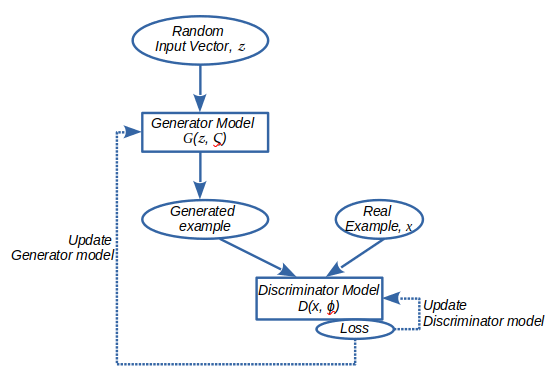}	
	\caption{General architecture of a GAN}\label{fig:gan}
\end{figure}

The general architecture of a GAN model is shown in Figure~\ref{fig:gan}. It is composed of two Neural Networks (NN): the generator and the discriminator. The generator model $G(z, \zeta$) is a network with parameters $\zeta$ that generates samples similar to real examples $x$ from a random vector $z$; the discriminator model is a network with parameters $\phi$ that seeks to classify samples (those artificial generated by the generator model and the real ones $x$) as real or fake. The two models are trained together using SGD: the discriminator uses the loss function in Eq.~\ref{eq:GAN_D_loss} to get better at discriminating real and fake samples, and the generator uses the loss function in Eq~\ref{eq:GAN_G_loss} to learn to produce examples able to fool the discriminator. Once this two-player game has converged (when the discriminator model obtains a probability close to 0.5 for each fake/real class), samples can be generated by $G(Z, \theta)$ from random vectors $z$.

\begin{equation}\label{eq:GAN_D_loss}
	\max_{\phi} \left[\mathbb{E}_{x}[\log D(x,\phi)] - \mathbb{E}_{z}[\log D(G(z,\theta),\phi)]\right]
\end{equation}

\begin{equation}\label{eq:GAN_G_loss}
\min_{\theta} \left[-\mathbb{E}_{z}\left[\log D(G(z,\theta),\phi) \right]\right], 
\end{equation}

\textbf{\textit{DP-GAN}}~\cite{DBLP:journals/corr/abs-1802-06739} applies DP-SGD method during the GAN training. The performance of DP-GAN is relatively poor in practice, but performance has been recently improving (see e.g. \cite{frigerio2019differentially}, \cite{Torkzadehmahani_2019_CVPR_Workshops}).

 An extension of PATE for training GAN models which is known as \textbf{\textit{PATE-GAN}}\footnote{See \url{https://github.com/BorealisAI/private-data-generation} for an open-source toolbox that implements PATE-GAN and several other methods for private data generation} was proposed in \cite{yoon2018pategan} and is described in Algorithm~\ref{alg:pategan}. The algorithms privately learn to discriminate fake from real labels using both private and generated data from the public generator. Then, teacher discriminators are ensembled to obtain the aggregate teacher; in turn, the aggregate teacher will use part of its private budget to predict the labels of a new public generated dataset. Consequently, the public generated dataset and the new private labelled dataset are merged and shuffled to be used to train the public discriminator and generator.  
So, the whole PATE-GAN algorithm works by alternatively generating datasets (with the generator model) and use them to a) train the aggregate teachers and b) update the discriminator by mixing them with privately labelled datasets. 

\noindent
\begin{minipage}{\textwidth}
\renewcommand*\footnoterule{}
\begin{algorithm}[H]
	\caption{PATE-GAN}\label{alg:pategan}
	
	\KwIn{
		\begin{itemize}\itemsep0em 
			\item Training data: $D \cup D_{\mathcal{G}}$ where $D_{\mathcal{G}}$ is a dataset generated by using the public generator model $\mathcal{M}_{\mathcal{G}}$
			\item Number of splits: $N$
			\item Private model: $\mathcal{M}_{pr}$
			\item Public GAN model with discriminatory model $\mathcal{M}_{\mathcal{D}}$; and generator model $\mathcal{M}_{\mathcal{G}}$
		\end{itemize}
	}
	\KwOut{Final parameters: $\theta$ of $\mathcal{M}_{\mathcal{G}}$ and $\phi$ of $\mathcal{M}_{\mathcal{D}}$}
	\vspace{0.3cm}
	Split randomly the dataset $\mathcal{D} \cup \mathcal{D_G}$ in $N$ parts, $\{\mathcal{D}_1,\dots, \mathcal{D}_N\}$
	
	\For {$i=1,\dots,N$}{
		Generate $f_i$ by training $\mathcal{M}_{pr}$ with the subset $\mathcal{D}_i$
	}
	
	Generate the aggregate teacher model, a noisy ensemble voting model:
	$f[x] = \underset{j}{\operatorname{argmax}} \left[ \sum_{i} [f_i[x]=j] + \xi \right]$, with $\xi \sim Lap[\epsilon^-1]$
	
	Produce a public dataset, $\mathcal{D'}$, using the generator model $M_\mathcal{G}$
	
	Produce $\mathcal{D''}$, a private generated dataset by using $f$ to label $\mathcal{D'}$
	
	Use $\mathcal{D'} \cup \mathcal{D''}$ to train $\mathcal{M}_{\mathcal{D}}$ and $\mathcal{M}_{\mathcal{G}}$ 
	
	\If {train has not ended\footnote{if $\mathcal{M}_{\mathcal{D}}$ and $\mathcal{M}_{\mathcal{G}}$ have converged or a maximum number of steps have been reached.}}{
		Generate another $D_{\mathcal{G}}$
		
		Go Step 1
	}
	
	Let $\theta$ and $\phi$ be the parameters of public GAN model.
	
\end{algorithm}
\end{minipage}

\begin{figure}[ht]
	\centering
	\includegraphics[width=\textwidth]{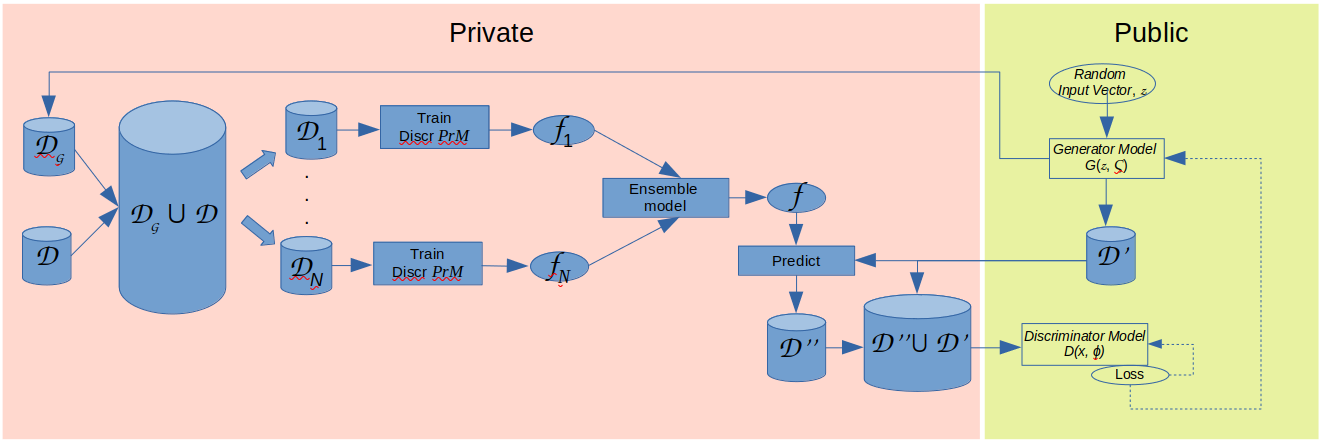}	
	\caption{General architecture and data processing of PATE-GAN algorithm}\label{fig:pategan}
\end{figure}

To evaluate the algorithm, the authors used the PATE-GAN algorithm to generate a dataset, and used this synthetic dataset to train a model, and assesses the quality if the PATE-GAN model using a real test dataset. PATE-GAN was significantly better than DP-GAN given the same privacy budget (although it was worse than training directly on the real data for some datasets), and only slightly worse than using the same method with a non-differentially private GAN. Therefore, it seems that the performance gap is caused by the GAN model itself and not from the DP mechanism. This constitutes an encouraging result in the area of DP and synthetic data generation. 

Whilst the majority of this work has been applied to image processing with high success~\cite{radford2015unsupervised}, ~\cite{DBLP:journals/corr/abs-1710-10196}, ~\cite{DBLP:journals/corr/abs-1809-11096}\footnote{See here: https://github.com/nashory/gans-awesome-applications for a curated list of GAN applications} less work has been done in Natural Language Processing (NLP). 

However, recent advances in this area are encouraging. Most of the work,~\cite{DBLP:journals/corr/YuZWY16, DBLP:journals/corr/CheLZHLSB17, Fedus2018, DBLP:journals/corr/abs-1804-11258, xu-etal-2018-diversity, ke2019araml} follows the proposal from~\cite{DBLP:journals/corr/YuZWY16} of mixing GANs with reinforcement learning, where the discriminator is trained to distinguish real from generated text samples and to provide rewards for the generator, which in turn is optimised via policy gradient. Rewards can contain also information about novelty and diversification of the texts such as in~\cite{xu-etal-2018-diversity}. The instability of the reinforcement learning on discrete data have been addressed in~\cite{ke2019araml} (by training the generator using maximum likelihood estimators and changing the discriminator rewards more those samples near the word distribution of real texts) and their system shown a significant improvement on with respect to the previous works. 

A new paradigm in~\cite{shao2019long} challenges the generation of long and diverse texts by the use of Variational Modelling. Their proposal follows the mechanisms of natural writing: first, planing a script and then filling it up with sentences that are generated upon the resulting script and to consider the previously generated context. This proposal not only outperforms state-of-the-art baselines for long texts but also provides an interesting efficient approach, ''decomposing long text generation into dependent sentence generation sub-tasks``.

\section{Local differential privacy}\label{sec:localdp}

One problem with differential privacy is that even if the statistics obtained over datasets are noised by a trusted server to protect users details, the real values are still saved in the datasets themselves, and therefore users have to trust the database maintainer to keep their privacy \cite{bebensee_1907-11908}.

%
%

A stronger privacy guarantee can be offered for individual users in the local differential privacy (LDP) setting where: 1) there is no need of a centralized authority and 2) each user can encode and perturb their inputs before transmitting them to the unstrusted server. The challenge with this setting is related to managing the overall variance of the perturbed input data the server receives, that in fact depends on the number of participants. So, most of the work on LDP has being applied to counting problems (training models in a LDP fashion is being addressed as Federated Learning with untrusted server as described in Section~\ref{sec:dpFL}):

\begin{description}[labelsep=0pt]
	\item[Frequency Oracles]: Estimation of locally private frequency, i.e. given a domain $\mathcal{D}$, a frequency oracle protocol estimates the frequency of an element $d \in \mathcal{D}$. For example, we could collect how many females are in a dataset without revealing the gender of a specific user. An abstract frequency oracle protocol and multiple specific implementations on top of it are described in \cite{Wang_10.5555/3241189.3241247}, which also contains several improvements in accuracy over well-known protocols as RAPPOR from Google.   
	
	\item[Heavy hitting identification]: In the case of large domains, instead of estimating the frequency of all of the elements $d \in \mathcal{D}$, we could estimate the frequency of the common domain elements (the heavy hitters). Efficient algorithms, such as \textit{TreeHist} and \textit{Bitstogram} are described in \cite{DBLP:journals/corr/BassilyNST17}.
	
	\item[Itemset mining]: Instead of computing frequency of single-valued inputs, this problem deals with the itemset mining in a local differential private settings. Specific algorithms can be found in \cite{10.1145/2976749.2978409} and \cite{8418600}.
	
	\item[Private spatial data collection]: Estimate the spatial distribution of users without revealing/saving the specific location of each user at a given time. A framework for this problem has been proposed in \cite{7498248}.
\end{description}

A few libraries implementing these types of LDP problems are available at: \url{https://github.com/vvv214/LDP_Protocols} (all of the above methods), \url{https://github.com/Samuel-Maddock/pure-LDP} (for frequency oracle and heavy hitting) and  \url{https://github.com/sisaman/LDP-mechanisms/blob/main/mechanisms.py} (for general types of mechanisms). 

How to combine LDP with the classic centralized DP as well as the application of ML on the top of LDP (and Deep Learning in particular) are still open questions, however a few steps towards these directions have been provided, especially in a federated context \cite{DBLP:journals/corr/abs-2009-03561}.

\section{Privacy in Federated Machine Learning}\label{sec:dpFL}

Traditional ML uses the centralized approach of sending a full dataset to the datacenter or machine where the learning will take place. In the Federated Learning (FL) setup, model owners send their models to the data holders, so the data never leaves the data holder premises, and a global model is trained from the parameters learned locally. This setup is more secure (and some authors classify FL as a privacy-preserving methodology), but still there are several types of attacks that can compromise data privacy (see section \ref{sec:fl_attacks} for more information about them). 

There are three main flavours of FL:

\begin{description}[labelsep=0pt]
	\item[Centralized FL]: Where there is a central server that orchestrate the different steps of the algorithm, coordinate all the participating nodes during the learning process and generates a global model. This setup can lead to a single point of failure in the aggregator side. To minimize privacy concerns local models are only sent to a trustable aggregator, in a framework known as Trustable FL (TFL). 
	\item[Decentralized FL]: Where the nodes are able to coordinate themselves to obtain the global model. This setup prevents single point failures but orchestration is more difficult and performance may be affected by the network topology. Rollout and blockchain-based federated learning frameworks are examples of decentralised FL.
	\item[Heterogeneous FL]: where there are no assumptions about the independence and distribution of the data, types of devices, collaborative schemes and models interacting in the learning, but all these resources are optimized to produce a single global model. Please, refer to \cite{DBLP:journals/corr/abs-2008-06767} extended explanation and examples.
\end{description}

In the following section we describe the most important types of attacks to FL systems and in sections \ref{sec:cfl} and \ref{sec:dfl}, we will describe practical examples of privacy solutions in the centralized and decentralized FL using DP\footnote{Other privacy-preserving techniques described in the literature such as Secure Multiparty Computation (SMPC) and Homomorphic Encryption (HE) are out of the scope of this review}. 

\subsection{Attacks in Federated Learning}\label{sec:fl_attacks}

(Mainly taken from \cite{2103-15753})

Attackers in FL aim to identify the underlining training data or trigger a miss-classification on the final model. Many of them have been identified and coined~\cite{2103-15753, 9308910, RodriguezBarroso2022SurveyOF} as:

\begin{description}[labelsep=0pt]
	\item[Model Inversion]: The aim is to reconstruct the training data. A potential attacker with access to target labels and model query, can query the final trained model to obtain the classification scores and reconstruct the rest of the data.
	\item[Membership Inference]: It uses similar strategy as the Model Inverse attack to identify if some data were part of the training dataset.
	\item[Model Encoding]: The attacker with access to the model tries to identify training data. In white-box models, the attacker uses the explainability/memorization power of the model to reveal training data; in black-box models, the attacker overfits the model to leak part of the target labels.
	\item[Model stealing]: Malicious participants could create a second model that mimics the decisions of the original model, and therefore avoid fee payment to the original ML experts or even sell the model to third parties.  
	\item[Model poisoning]: Malicious participants submit a local model that triggers specific results given specific inputs, for example modifying a classifier that assigns an attacker-chosen label to certain features of the data.
	\item[Data poisoning]: Malicious participants introduces ``bad'' training data, making the model accuracy drop. 
	\item[Adversarial examples]: The attacker does not require access to the training procedure, it tries to trick the model in order to falsely classify a prediction. A potential attacking is a malware that evades the detection of a ML intrusion detection system.
\end{description}


The three most important techniques for preventing these attacks are: 1) applying differential privacy to the parameters of the local models, so they do not leak any personal data; 2) using Secure Multiparty Computation (SMPC), in which gradients and parameters are cryptographically split amongst the participants of the computation and all the data processing is done in a decentralised manner; 3) Homomorphic Encryption (HE), in which the ML training is done over encrypted data.

Whilst DP is a non-expensive technique, it produces a decrease in accuracy; SMPC does not protect during testing phase of ML; and HE is a promising method but with a very high computational cost which is not tolerable in real-world situations.

Many other defence methods are available in the literature (see \cite{2103-15753} for a comprehensive list of them), but unlike DP, SNPC and HE, they are tailored to specific types of attacks and therefore less useful in a practical and more broad setups.

\subsection{Privacy in Centralized FL using DP}\label{sec:cfl}

Most works in centralized FL with DP are focused on obfuscating gradient variations on each of the clients. Here we discussed three alternatives described in the literature:

\begin{description}[labelsep=0pt]
	\item[Random subsampling and distorsion]: In \cite{1712-07557} the authors propose a DP mechanism in which DP is applied on the aggregator side by using a randomised mechanism and distortion of the updates. The random subsampling consists of creating rounds that involve only a random subset of clients updating the model and submitting the updated gradient; distortion is done by using a Gaussian mechanism that clips the updates and add noise, both depending on the sensitivity of the updates. Their results show that at early communication rounds a small number of clients can contribute, but in later rounds the number of users involved needs to increase to gain accuracy. In setup with reduced number of users Gaussian mechanism cannot be applied (due to trade off between noise-utility) and also, random subsampling may not obfuscate the information about the involved clients in each round.
	
	\item[Random masks and adaptative quantization]: Parameter quantization has been  widely used for DNN model compression for both training and testing \cite{DBLP:journals/corr/abs-1712-01048}. In \cite{2109-05666} it is used to produce local differential privacy in the sense that each client (in this case, a sensor) transmit DP parameters. The DP-parameters can be obtained combining random masks with adaptative quantization to reduce the number of parameters sent to the aggregators, and improve the privacy leakage from the gradients. Random mask mechanism simply generates a sparse matrix from the gradient matrix, by replacing portion of random set of the gradients by zero. Adaptative quantization compresses each weight matrix in the DNN, $W$, based on one-bit adaptive quantization, that is $W_{i,j} \sim Poisson(p_{i,j})$, $p_{i,j}=\frac{W_{i,j}-min_{i,j} W}{max_{i,j} W - min_{i,j} W}$. This paper does not relates these heuristics with DP concepts, but it is very difficult to reconstruct a $W$ matrix from the quantized one and it can be considered as an alternative to DP mechanisms that at the same time solve communication problems for large models. Their results over a LSTM network are encouraging as a very similar accuracy is obtained using or not adaptative quantization.
	
	\item[Smashed data and gradient sign]: In \cite{ma2021federated}
	the training is divided in two phases. In a first phase, devices learn compact but informative representations of the raw data locally, that they call smashed data. In the second phase, the shared global model uses these smashed data as input, rather than the raw data. Parameters are not sent to the server, instead, a 1-bit sign of their gradient is sent. The server then aggregates the signs and send them back to the clients. The global model is aggregated locally on the server based on the global loss and the local loss sent from the clients. 

\end{description}


Privacy in Federated Learning can be further boosted by using LDP during training as done in~\cite{DBLP:journals/corr/abs-2106-09779}\footnote{
	\url{https://github.com/lowya/Locally-Differentially-Private Federated-Learning}} where the authors have trained a Logistic regression model in a Federated fashion noising the local models.

\subsection{Privacy in Decentralized FL using blockchain}\label{sec:dfl}

Recently, private decentralized FL has been addressed using blockchain technology. Two of the more interesting approaches are described in \cite{2007-03856} and \cite{2103-15753}:
\begin{itemize}	
	\item \textit{BlockFLow} is described in \cite{2007-03856} as an accountable federated learning system that is fully decentralized and privacy-preserving. It incorporates differential privacy, introduces a novel auditing mechanism for model contribution, and uses Ethereum smart contracts to incentivize good behaviour and keep all of the interactions during training. Their system does not require a centralized test dataset, sharing of datasets directly between the agents, or trusted auditors; it is fully decentralized and resilient up to a 50\% collusion attack in a malicious trust model. ``Each agent must run their own instance of the BlockFLow client, which handles all aspects of an agent’s participation in a federated learning experiment. For every federated learning round, each BlockFLow client trains a local model, applies differential privacy by adding Laplacian noise to the model, shares its model with the other clients in the experiment by using IPFS, retrieves and evaluates other clients’ models, reports the evaluation scores to the BlockFLow smart contract, and retrieves the overall scores and averages the clients’ models.''. Contribution scoring procedure described in the paper penalizes contributions with malicious models, promotes the submission of strong contributor models honesty during evaluation. The major drawback of this solution is the overhead time due to all of the cryptographic operations involved during the training process (when a data/model is retrieved or saved to IPFS or during the evaluation of other agent models).
	
	\item In \cite{2103-15753} the authors use the \textit{DIDComm messaging} implementation available in Hyperledger Aries, a toolkit designed ``for initiatives and solutions focused on creating, transmitting and storing verifiable digital credentials''\footnote{https://www.hyperledger.org/use/aries} to create a secure communication channel to execute FL. 
	Hyperledger Aries blockchain maintains all the peer's DIDs and DID documents\footnote{https://www.w3.org/TR/did-core/} and emits credentials that can be used to execute the services described in the DID documents. In their proposal, ML practitioners, Regulators and organizations (their use case is around Medical Health Records and therefore organizations are mainly formed by NHS Trust agents) are connected to a Hyperledger Aries network and when a FL process is requested, their protocol allows all the peers to exchange and verify DIDs and establish trusted connections. Once all trusted connections are established, a roll up FL algorithm can be executed.
	
\end{itemize}

\section{Discussion}\label{sec:discussion}


The two more popular DP ML algorithms (DP-SGD and PATE) were discussed: the first one re-structures SGD to add noise to the gradients conveniently during training so model parameters do not leak any private information; the second completes a public unlabeled dataset with differentially private labels produced from a noised ensemble model and uses this dataset, in turn, to create a DP model.

Ideas of the PATE algorithm resembles synthetic data generation, and in fact, it is one of the approaches to avoid leaking private data during data processing. However, statistical approaches for synthetic data generation have proven to be very difficult and GAN models can be used to produce better results. Merging the ideas of these two algorithms, PATE-GAN is intended to produce synthetic data by integrating a PATE-kind discriminator to generate DP data which in turns can be use to produce a DP GAN model.

Local Differential Privacy can be used in scenarios where the original sensitive data do not need and/or can not be centrally stored and a DP version of these data can be used instead. LDP applications are very difficult to configure and optimise due to the disparities of the noise added at the different locations to preserve individualities. So, applications heavily relying on counters are typical use cases of LDP, but LDP has been also used in federated contexts to avoid sharing the real gradients or parameters with other parties in both centralized and decentralized frameworks. 

Solutions in decentralised frameworks are mainly focused on the integration of DP with a blockchain technology. This symbiosis  implies a high latency during processing; however, it brings, amongst other advantages, a high level of privacy guarantee as the operational traceability in blockchain prevents any attack without revealing the attacker identity and therefore prevents the attack happening in the first place. 

\begin{figure}[ht]
	\centering
	\includegraphics[width=0.8\textwidth]{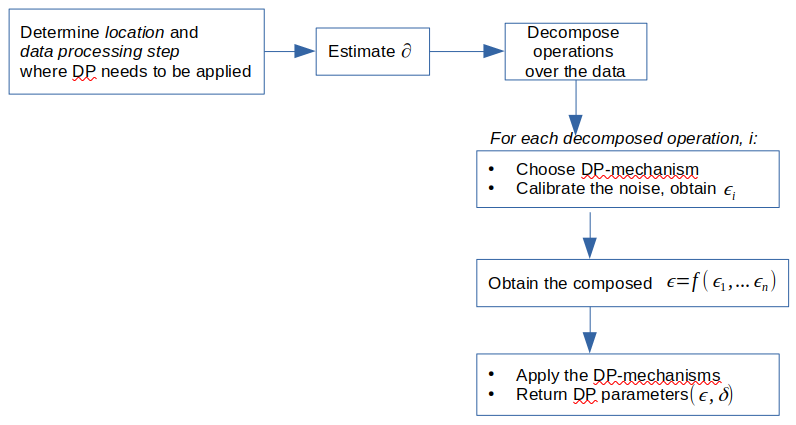}	
	\caption{Guideline for applying DP during in data processing.}\label{fig:overview}
\end{figure}

Figure~\ref{fig:overview} offers a guideline for applying DP in practice. The first step is to determine which data and data processes are susceptible to having privacy issues. The data practitioner should consider that in the local vs centralised setup, the original, sensitive data should not be centrally saved, and that the noise applied to the data at each location could be different and therefore affect the utility of the overall processing. As DP could be applied at different stages of data processing, the data used in a DP-mechanism will be DP-protected as long as 1) only the outcome of a DP process is revealed along with the DP parameters and 2) downstream processes uses the outcome of this DP process without accessing the original source. 

The next step is to chose a maximum allowed $\delta$ (risk probability due to uncontrolled causes). Data processing is next decomposed so that a DP-mechanism can be applied to each component. The DP-mechanisms used should be chosen according to their suitability for the task. Then, the final composed value of $epsilon$ should be computed and checked against the privacy agreements with the users. Finally, the selected DP-mechanisms can be applied, and the results of the processing returned along with the DP-parameters $(\epsilon, \delta)$.

To enhance privacy the full data processing described in Figure~\ref{fig:overview} could be embedded in a blockchain application and all privacy parameters along with metrics to encourage collaboration and honest behaviour could be, consequently, saved on a blockchain.

\section{Conclusions}\label{sec:conclusions}

This paper has summarised the most important concepts about Differential Privacy. First, a deep understanding of the DP definition and their variants was presented. We highlighted the importance of variants as Reny-DP and f-DP, specially for ML applications, and listed the libraries where available implementations can be found. 

The relation of DP with synthetic data generation, in particular with GAN methods, as well as the difference between local and centralised DP were all described here and relevant works were detailed. 

Finally, we offer the reader a practical guideline for applying DP.

\section{Acknowledgements}

The author would like to thank Dave Lewis, Clinton Swan and Rahul Sharma whose useful review and comments helped to significantly improve this paper’s presentation.

\bibliographystyle{unsrt}
\bibliography{dp}

\end{sloppypar}
\end{document}